\documentclass[12pt]{article}
 \usepackage{amsmath,amssymb,amsbsy,amstext,amscd,amsfonts}
 \usepackage{exscale,calrsfs,times,multind,theorem}
 \usepackage{color}
 \usepackage{pictex,epsfig}

\newtheorem{lemma}{\indent Lemma}[section]
\newtheorem{theorem}[lemma]{\indent Theorem}

\newtheorem{proposition}[lemma]{\indent Proposition}

\newtheorem{definition}[lemma]{\indent Definition}

\newtheorem{remark}[lemma]{\indent Remark}

\newcommand{\al}{\alpha}

\newcommand{\Gm}{\Gamma}

\newcommand{\vf}{\varphi}
\newcommand{\ve}{\varepsilon}

\newcommand{\cO}{{\mathcal O}}
\newcommand{\bB}{\mathbb{B}}
\newcommand{\bC}{\mathbb{C}}
\newcommand{\bH}{\mathbb{H}}
\newcommand{\bL}{\mathbb{L}}
\newcommand{\bR}{\mathbb{R}}
\newcommand{\bS}{\mathbb{S}}
\newcommand{\bW}{\mathbb{W}}

\newcommand{\nub}{{\boldsymbol{\nu}}}

\newcommand{\cC}{{\mathcal C}}

\newcommand{\cF}{{\mathcal F}}

\newcommand{\cH}{{\mathcal H}}
\newcommand{\cK}{{\mathcal K}}

\newcommand{\cM}{{\mathcal M}}

\newcommand{\cS}{{\mathcal S}}

\newcommand{\cx}{{}{\scriptstyle{\mathcal X}}} 

\newcommand{\dst}{\displaystyle}
\newcommand{\pa}{\partial}
\newcommand{\ov}{\overline}
\newcommand{\wt}{\widetilde}

\newcommand{\A}{{\boldsymbol A}}

\newcommand{\F}{{\boldsymbol F}}

\newcommand{\J}{{\boldsymbol J}}
\newcommand{\K}{{\boldsymbol K}}
\newcommand{\M}{{\boldsymbol M}}
\newcommand{\N}{{\boldsymbol N}}

\newcommand{\T}{{\boldsymbol T}}

\newcommand{\V}{{\boldsymbol V}}
\newcommand{\W}{{\boldsymbol W}}

\newcommand{\supp}{\operatorname{supp}}
\newcommand{\QED}{\hspace{\fill}$\Box$\medskip\par}

\title{Mixed boundary value problems for the Helmholtz equation in a model 2D angular domain}

\author{R. Duduchava, \& M. Tsaava}
 
\begin{document}
 \date{}
 \maketitle

\begin{abstract}
The purpose of the present research is to investigate model mixed boundary value problems for the Helmholtz equation in a planar angular domain $\Omega_\alpha\subset\mathbb{R}^2$ of magnitude $\alpha$. The BVP is considered in a non-classical setting
when a solution is sought in the Bessel potential spaces $\mathbb{H}^s_p(\Omega_\alpha)$, $s>1/p$, $1<p<\infty$. The problems are investigated using the potential method by reducing them to an equivalent boun\-dary integral equation (BIE) in the Sobolev-Slobode\v{c}kii space on a semi-infinite axes $\bW^{s-1/p}_p(\bR^+)$, which is of Mellin convolution type. By applying the recent results on Mellin convolution equations in the Bessel potential spaces obtained by V. Didenko \& R. Duduchava in \cite{DD16}, explicit conditions of the unique solvability of this BIE in the Sobolev-Slobode\v{c}kii $\bW^r_p(\bR^+)$ and Bessel potential $\bH^r_p(\mathbb{R}^+)$ spaces for arbitrary $r$ are found and used to write explicit conditions for the Fredhoilm property and unique solvability  of the initial model BVPs for the Helmholtz equation in the above mentioned non-classical setting.

The same problem was investigated in the foregoing paper \cite{DT13}, but there was made fatal errors. In the present paper we correct these results.
\end{abstract}



{\bf ﻿Funding:} The research was supported by Shota Rustaveli National Science Foundation grants no. 13/14 and 31/39 and completed during the visit of the first author in the Saarlands University, Saarbr\"ucken, Germany in 2016, under the funding of Humboldt Foundation.


\maketitle


\section*{Introduction and formulation of the main results}
\addcontentsline{toc}{section}{Introduction and formulation of the problems}
\label{section0}

Consider the model domain $\Omega_\alpha$ which is the plane angle of magnitude $\alpha$ between the half axes $\bR^+$ and the beam $\bR_\alpha$ turned by the angle $\alpha$ from $\bR^+$ (see Fig. 1). The corresponding boundary is a model curve:
\begin{eqnarray}\label{e0.1}
\begin{array}{c}
\Gamma_\alpha:=\pa\Omega_\alpha=\bR^+\cup\bR_\alpha,\qquad \bR^+=[0,\infty),\qquad 0<\alpha<2\pi,\\[2mm]
     \bR_\alpha:=\{e^{i\alpha}t=(t\cos\,\alpha,t\sin\,\alpha)\;:\; t\in\bR^+\}.
\end{array}
 \end{eqnarray}
Note, that the case $\alpha=\pi$ is already well treated in the literature.

The unit normal vector field $\{\nub(x)\}_{x\in\Gamma_\alpha}$ on the boundary $\Gamma_\alpha$ is defined by the equality
\begin{eqnarray}\label{e0.2}
\nub(x)=\left\{\begin{array}{ll}
  (0,-1)^\top & \hbox{\rm for}\quad x\in\bR^+\\[3mm]
  (-\sin\,\alpha,\cos\,\alpha)& \hbox{\rm for}\quad x\in\bR_\alpha.\end{array}\right.
 \end{eqnarray}
and defines  the following normal derivative $\pa_{\nub}$ on the boundary:
\begin{eqnarray}\label{e0.3}
\partial_{\nub(t)}=\left\{\begin{array}{ll}
  - \dst\lim_{(x_1,x_2)\to{t=(\tau,0)}}\pa_{x_2} & \text{for}\;\; t\in \bR^+,\\[3mm]
  \dst\lim_{(x_1,x_2)\to t=(\tau\cos\,\alpha,\tau\sin\,\alpha)}[-\sin\, \alpha\,\pa_{x_1}+\cos\,\alpha\,\pa_{x_2}] & \text{for}\;\;t\in\bR_\alpha. \end{array}\right.
 \end{eqnarray}

\setlength{\unitlength}{0.4mm}
\hskip-3mm
\begin{picture}(300,140)
\put(0,-20){\epsfig{file=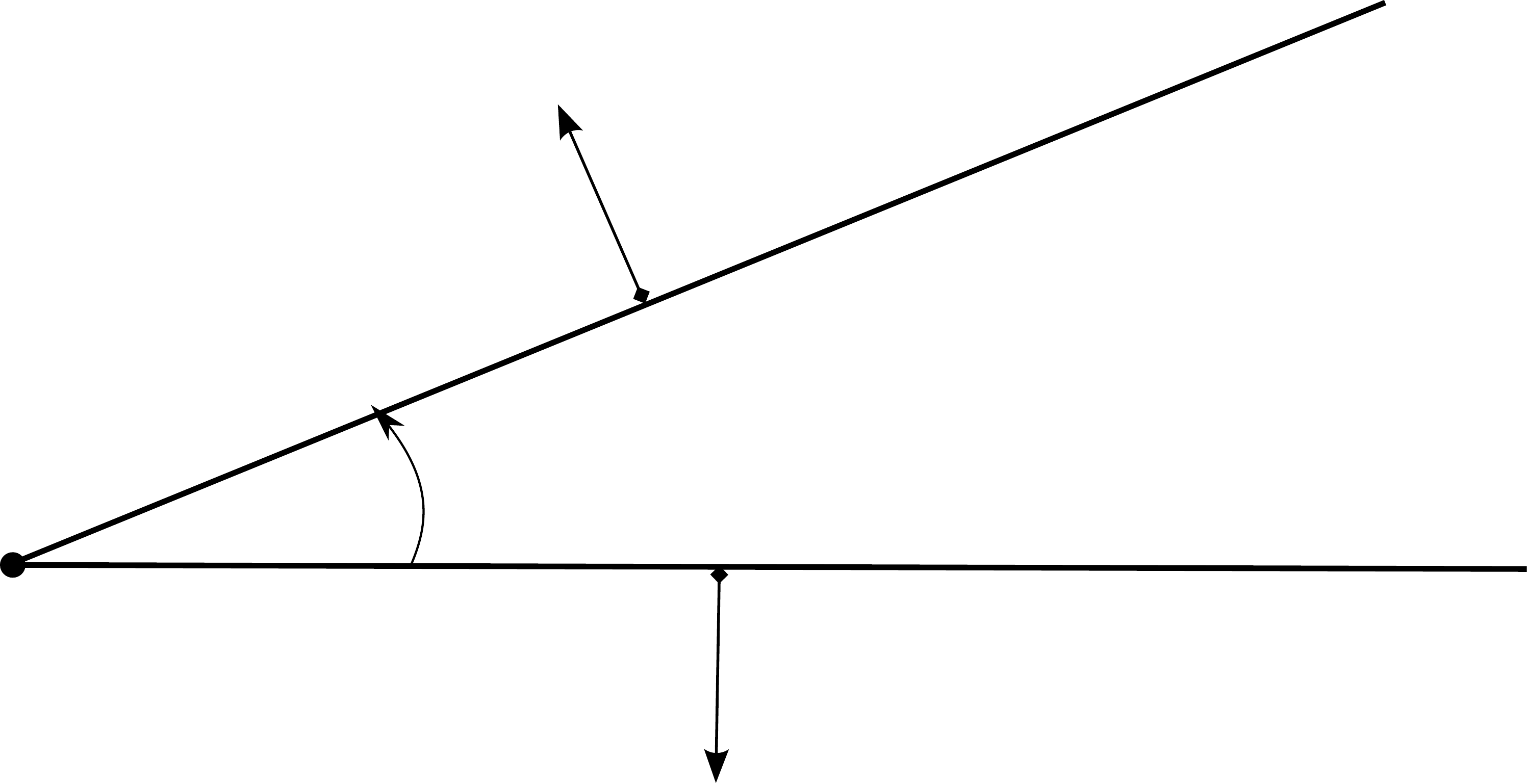,height=60mm, width=100mm}}
\put(0,30){\makebox(0,0)[lc]{$0$}}
\put(155,60){\makebox(0,0)[bc]{$t=(\tau\,\cos\,\alpha,\tau\,\sin\,\alpha)^\top$}}
\put(110,110){\makebox(0,0)[bc]{$\nub(t)=(-\sin\,\al,\cos\,\al)^\top$}}
\put(140,22){\makebox(0,0)[bc]{$t=(\tau,0)^\top$}}
\put(162,-15){\makebox(0,0)[bc]{$\nub(t)=(0,-1)^\top$}}
\put(83,35){\makebox(0,0)[bc]{$\alpha$}}
\put(150,37){\makebox(0,0)[bc]{$\Omega_\alpha$}}
\put(230,23){\makebox(0,0)[bc]{$\bR^+$}}
\put(200,121){\makebox(0,0)[bc]{$\bR_\alpha$}}
\end{picture}
\vskip-3mm
\hskip95mm{\rm Fig. 1}
 \vskip9mm

Many problems in mathematical physics e.g., cracks in elastic media, electromagnetic scattering by surfaces etc., are formulated in the form of boundary value problems for elliptic partial differential equations in domains with angular points at the boundary. In the recent paper \cite{BDKT13} is described how such BVPs can be investigated with the help of their local representatives-model problems in planar angles
$\Omega_{\alpha_j}$ of magnitude $0<\alpha_j<2\pi$, $j=1,\ldots,m$. The purpose of the present paper is to study the model mixed boundary value problem for the Helmholtz equation in the model domain $\Omega _\alpha$
\begin{eqnarray}\label{e0.4}
\left\{\begin{array}{ll}
\Delta u(x)+k^2u(x)=f(x),\qquad & x\in\Omega_\alpha, \\[0.2cm]
u^+(t)=g(t),     \qquad & t\in\mathbb{R}_\alpha, \\[0.2cm]
(\partial_\nub u)^+(t)=h(t),\qquad & t\in\mathbb{R}^+,
\end{array}\right.
\end{eqnarray}
for a complex parameter ${\rm Im}\,k\not=0$ Here $u^+$ and $(\partial_{\nub_\Gamma}u)^+$ denote respectively the Dirichlet and the Neumann traces on the boundary.

Let us recall short definitions of the Bessel potential $\mathbb{H}^s_p(\Omega_\alpha)$, $\widetilde{\mathbb{H}}^s_p( \Omega_\alpha)$, $\mathbb{H}^r_p(\bR^+)$, $\widetilde{\mathbb{H}}^r_p(\bR^+)$ and Sobolev-Slobode\v{c}kii $\widetilde{\mathbb{W}}^r_p(\bR^+)$, $\mathbb{W}^r_p(\bR_\alpha)$ etc. spaces for $r\in\bR$, $1<p<\infty$. The spaces  $\mathbb{H}^r_p(\Gamma_\alpha)$, $\widetilde{\mathbb{H}}^r_p(\Gamma_\alpha)$, $\mathbb{W}^r_p(\Gamma_\alpha)$ and $\widetilde{\mathbb{W}}^r_p(\Gamma_\alpha)$ can only be defined for $-2+1/p<r<1+1/p$. For detailed definitions and properties of these spaces we refer to the classical source \cite{Tr95} and also to \cite{CD01,Du84a,Du01,DS93,Hr85,HW08}.

Bessel potential space $\bH^s_p(\bR^n)$ is defined as a subset of the space of Schwartz distributions $\bS'(\bR^n)$ and is endowed with the following norm (see \cite{Tr95}):
 \[
  ||u\big|\bH_p^s(\bR^n)||:=||\langle D\rangle^su\big| L_p(\bR^n)||,\quad
  \mbox{ where }\quad \langle D\rangle^s:=\cF^{-1}(1+|\xi|^2)^{\frac s2}\cF.
 \]
$\cF$, $\cF^{-1}$ are the Fourier transforms.  For the definition of the Sobolev-Slobode\v{c}kii space $\bW_p^s(\bR^n)=\bB_{p,p}^s(\bR^n)$ see \cite{Tr95}.

The space $\wt {\bH}_p^s(\Omega)$,  where $\Omega\subset\bR^n$,  is defined as the subspace of $\bH_p^s(\bR^n)$ of those functions (or distributions if $s<0$) $\vf\in \bH_p^s(\bR^n)$, which are supported in the subset $\Omega$, $\supp\vf\subset\ov{\Omega}$, whereas $\bH_p^s(\Omega)$ denotes the quotient space $\bH_p^s(\Omega):=\bH_p^s(\bR^n)\Big/\wt{\bH}_p^s(\Omega^c)$ and $\Omega^c:=\bR^n\setminus\ov{\Omega}$ is the complemented domain. The space $\bH_p^s(\Omega)$  can be identified with the space of distributions $\vf$ on $\Omega$ which admit extensions $\ell\vf\in\bH_p^s(\bR^n)$. Therefore $r_\Omega\bH_p^s(\bR^n)=\bH_p^s(\Omega)$, where $r_\Omega$ denotes the restriction  from $\bR^n$ to  the domain $\Omega$.

Worth noting that for an integer $m=1,2,\ldots$ the spaces $\bH^m_p(\bR^n)$ and $\bW^m_p(\bR^n)$ coincide and are known as the Sobolev spaces, endowed with the following  equivalent norm (see \cite{Tr95}):
 \[
  ||u\big|\bW_p^m(\bR^n)||:=\sum_{|\alpha|\leqslant m}||\pa^\alpha u\big| L_p(\bR^n)||.
 \]

 Let $\cS:=\partial\Omega$ be the smooth boundary and consider $\widetilde{\mathbb{H}}^{-1}_0(\Omega)$, a subspace of $\widetilde{\mathbb{H}}^{-1}(\Omega)$, orthogonal to
 \[
\widetilde{\mathbb{H}}^{-1}_\cS (\Omega):=\left\{f \in\widetilde{\bH}^{-1} (\Omega)\;:\;\langle f,\varphi\rangle = 0 \;\text{for all}\;\varphi\in C^1_0(\Omega)\right\},
 \]
where $\langle\cdot,\cdot\rangle$ denotes the pairing between the adjoint spaces and coincides with the usual scalar product for regular functions.
$\widetilde{\mathbb{H}}^{-1}_\cS(\Omega)$ consists of those distributions on $\Omega$, belonging to  $\widetilde{\mathbb{H}}^{-1}(\Omega)$ which have their supports just on $\cS$ and  $\widetilde{\mathbb{H}}^{-1} (\Omega)$ can be decomposed into the direct sum of subspaces which are orthogonal to each-other:
  \[
 \widetilde{\mathbb{H}}^{-1}(\Omega)= \widetilde{\mathbb{H}}^{-1}_\cS(\Omega)\oplus
       \widetilde{\mathbb{H}}^{-1}_0(\Omega).
  \]
 The space $\widetilde{\bH}^{-1}_\cS(\Omega)$ is non-empty (see \cite[\S\, 5.1]{HW08}) and excluding it from $\widetilde{\mathbb{H}}^{-1}(\Omega)$ is necessary to make  BVPs uniquelly solvable (cf. \cite{HW08} and the next Theorem \ref{t0.1}).

Let us commence with the existence results for a solution to BVP \eqref{e0.4}. Lax-Milgram Lemma applied to these BVPs gives the following solvability result.
 %
\begin{theorem}\label{t0.1}
The Mixed BVP \eqref{e0.4} has a unique solution in the classical weak setting:
\begin{equation}\label{e0.5}
u\in\bH^1(\Omega_\alpha),\qquad f\in\widetilde{\mathbb{H}}^{-1}_0(\Omega_\alpha),\qquad g\in\mathbb{H}^{1/2}(\bR_\alpha),\qquad h\in\mathbb{H}^{-1/2}(\bR^+).
\end{equation}
\end{theorem}
{\bf Proof:} The proof is verbatim to the proof of similar Theorem 2.1 (also see Remark 2.2 and Remark 2.3) in \cite{DTT14}. The operator treated in \cite{DTT14}, is very similar: Laplace-Beltarmi with $0$ order summand on a compact surface. The main tool,the Lax-Milgram Lemma applies equally successful for non-compact domain.        \QED

As we see from Theorem \ref{t0.1} the BVP \eqref{e0.4} has a unique solution in the classical setting \eqref{e0.5} independent of the values of angular points on the boundary. This property changes dramatically as soon as we consider the BVP \eqref{e0.4} in the following non-classical setting
\begin{eqnarray}\label{e0.6}
\begin{array}{r}
u\in\mathbb{H}^s_p(\Omega_\alpha),\quad f\in\widetilde{\mathbb{H}}^{s-2}_p(\Omega_\alpha)\cap
     \widetilde{\mathbb{H}}^{-1}_0(\Omega_\alpha),\quad g\in\mathbb{W}^{s-1/p}_p(\bR_\alpha),\\[3mm]
h\in\mathbb{W}^{s-1-1/p}_p(\bR^+), \qquad 1<p<\infty, \quad
    \dst\frac1p<s<1+\dst\frac1p.
\end{array}
\end{eqnarray}
(see Remark \ref{r0.4} below). This indicates that the derivative of a solution has singularities depending on the angles on the boundary; although this is knows long ago, we find for the first time all values of "forbidden angles" in Theorem \ref{t0.3} below.

Note that, the upper constraint in  $\dst\frac1p<s<1+\dst\frac1p$ ensures the invariant definition of the Bessel potential and Sobolev-Slobode\v{c}kii spaces, while the lower constraint ensures the existence of the trace $u^+$ on the boundary.

Moreover, from Theorem \ref{t0.1} we can not even conclude that a solution is continuous in the closed domain $\overline{\Omega}_\alpha$. If we can prove that there is a solution $u\in\mathbb{H}^1_p(\Omega_\alpha)$ for some $2<p<\infty$, we can enjoy even a H\"older continuity of $u$ in $\overline{\Omega}_\alpha$. It is very important to know maximal smoothness of a solution in some problems, for example in approximation methods.

Along with the non-classical settings \eqref{e0.6} the trace of a solution $u^+$ on the boundary $\Gamma_\alpha$ satisfies the following compatibility conditions:
\begin{eqnarray}\label{e0.7}
\begin{array}{c}
u^+_+ - \J_\alpha u^+_\alpha\in\wt{\bW}^{s-1/p}_p(\bR^+),\\[3mm]
     (\pa_\nub u)^+_+ +\J_\alpha (\pa_\nub u)^+_\alpha\in\wt{\bW}^{s-1/p-1}_p(\bR^+),
\end{array}
\end{eqnarray}
where $v^+_+$ denotes the trace on $\bR^+$, $v^+_\alpha$ denotes the trace on $\bR_\alpha$ and
 \begin{eqnarray}\label{e0.8}
\J_\alpha\vf(t)=\vf(t\,\cos\,\alpha,t\,\sin\,\alpha),\qquad  t\in\bR^+
 \end{eqnarray}
is the pull back operator from $\bR_\alpha$ to $\bR^+$. These compatibility conditions are direct consequences of the inclusion
$u\in\bH^s_p(\Omega_\alpha)$ and the properties of traces on the boundary
$\Omega_\alpha$.

To formulate the main theorem of the present work we need the following definition.
 %
\begin{definition}\label{d0.2}
The BVP \eqref{e0.4}, \eqref{e0.5}  is Fredholm if the homogeneous problem $f=g=h=0$ has a finite number of linearly independent solutions and the BVP has a solution if and only if the data $f,g,h$ satisfy a finite number of orthogonality conditions.
\end{definition}

Next we formulate the main theorem of the present paper, which is proved in \S\, \ref{sect5}.
 %
\begin{theorem}\label{t0.3}
Let $\alpha\in(0,2\pi)$, $1<p<\infty$ and $\dst\frac1p<s<1+\dst\frac1p$.  The model mixed BVP \eqref{e0.4} is Fredholm in the non-classical setting \eqref{e0.6} if and only if:
 \begin{eqnarray}\label{e0.10}
e^{4\pi(s-1/p)i}\sin^2\pi\left(s-\dst\frac2p\right)+\cos^2(\pi -\alpha)\left(s-\dst\frac2p\right)\not=0.
 \end{eqnarray}
In details the condition \eqref{e0.10} is written as follows:
{
\begin{subequations}
 \begin{eqnarray}\label{e0.10a}
&&\hskip-17mm 1)\;\; s\not=\dst\frac1p+\dst\frac n4\quad n=1,2,3\quad
     \text{and}\quad s\not=\dst\frac2p+n,\quad n=0,-1,\\
\label{e0.10b}
&&\hskip-17mm 2)\;\;\text{If}\quad s=\dst\frac2p-1,
     \quad\text{then}\quad\alpha\not=\frac\pi2,\,\frac{3\pi}2,\\
\label{e0.10c}
&&\hskip-17mm 3)\;\;\text{If}\;\;s=\frac1p+\dst\frac12,\;\;
     \text{then}\;\; p\not=\dst\frac2{n-2m}\;\; \text{or}\;\;\alpha\not=\dst\frac{2k+1}{2m}\pi,\\
\label{e0.10d}
&&\hskip-19mm \begin{array}{l}
4)\;\;\text{If}\quad s=\dst\frac1p+\dst\frac n2+\dst\frac14,\quad n=0,1, \quad\text{then}\\
     \quad\; \alpha\not=2\pi p\dst\frac{2k+1}{2np+p-4}\quad \text{and}
     \quad \alpha\not=4\pi\dst\frac{np-kp-2}{2np+p-4},
     \end{array}
 \end{eqnarray}
\end{subequations}}
where $k,m=0,\pm1,\ldots.$

In particular, the BVP \eqref{e0.4} has a unique solution in the settings \eqref{e0.6} if:
{
 \begin{eqnarray}
\begin{array}{l} \label{e0.11}
\dst\frac1p+\dst\frac14<s\leqslant\dst\frac2p\quad\text{for}\quad
     1<p\leqslant2,\\[3mm]
\dst\frac2p\leqslant s<\dst\frac1p+\dst\frac34,\quad\text{for}\quad
     2\leqslant p<\infty.
 \end{array}
 \end{eqnarray}}
\end{theorem}
 %
\begin{remark}\label{r0.4}
From \eqref{e0.11} follows directly that the BVP \eqref{e0.4} has a unique solution in the setting \eqref{e0.6} but for $p=2$ if
 \begin{eqnarray}\label{e0.11a}
\dst\frac34<s<\dst\frac54.
 \end{eqnarray}

As we already know from Theorem \ref{e0.1} and as it follows from \eqref{e0.11a}, for $p=2$, $s=1$ (the classical setting) the BVP \eqref{e0.4} has no "forbidden angles" $\alpha$ and the problem is uniquely  solvable for all values of $\alpha$ (cf. Theorem \ref{t0.1}). {The "forbidden angles" $\alpha$ might only emerge when $p\not=2$ or $s\not=1$ and for these angles the problem is not even Fredholm.}
\end{remark}

Theorem \ref{t0.3} is a corollary of the next two Theorem \ref{t0.5} and Theorem \ref{t0.6}.
 %
\begin{theorem}\label{t0.5}
Let $1<p<\infty$ and $\dst\frac1p<s<1+\dst\frac1p$. Let $g_0\in\bW^{s-1/p}_p(\Gamma_\alpha)$ and $h_0\in\bW^{s-1-1/p}_p(\Gamma_\alpha)$ be some fixed extensions of the boundary conditions $g\in\bW^{s-1/p}_p(\bR_\alpha)$ and $h\in\bW^{s-1-1/p}_p(\bR^+)$ (non-classical case),  defined initially
on parts of $\Gamma_\alpha$.

A solution to the BVP \eqref{e0.4} is represented by the formula
\begin{eqnarray}\label{e0.12}
u(\cx)=\N_\cC f(\cx)+\W_\Gamma(g_0+\varphi_0)(\cx)-\V_\Gamma(h_0
     +\psi_0)(x),\qquad x\in\Omega_\alpha,
 \end{eqnarray}
where  $\varphi_0$ and $\psi_0$ are solutions to the following
system of pseudodifferential equations
\begin{eqnarray}\label{e0.13}
&&\left\{\begin{array}{ll}
\dst\frac12\varphi_0(t)+r_{\bR^+}\V_{\Delta+k^2,-1}\psi_0(t)
     =G_+(t),\qquad &  t\in\bR^+,\\[3mm]
\dst\frac12\psi_0(t) - r_{\bR_\alpha}\V_{\Delta+k^2,+1}\varphi_0(t)
     =H_-(t)\qquad & t\in\bR_\alpha, \end{array}\right. \\[2mm]
&&\hskip15mm\varphi_0\in\wt{\bW}^{s-1/p}_p(\bR^+), \qquad
     \psi_0\in\wt{\bW}^{s-1-1/p}_p(\bR_\alpha),\nonumber\\[2mm]
&&\hskip15mm G_+:=r_{\bR^+}G_0\in\bW^{s-1/p}_p(\bR^+),\qquad
     H_-=r_{\bR_\alpha}H_0\in\bW^{s-1-1/p}_p(\bR_\alpha),\nonumber\\
&& G_0:=(\N_{\Delta+k^2}f)^+-\dst\frac12g_0+\W_{\Delta+k^2,0}g_0
     -\V_{\Delta+k^2,-1}h_0\in\bW^{s-1/p}_p(\Gamma_\alpha),\nonumber \\
&& H_0:=(\pa_\nub\N_{\Delta+k^2} f)^+-\dst\frac12h_0
     +\V_{\Delta+k^2,+1}g_0-\W^*_{\Delta+k^2,0}h_0\in\bW^{s-1-1/p}_p
     (\Gamma_\alpha).\nonumber
\end{eqnarray}
Vice versa: if $u$  is a solution to the BVP  \eqref{e0.4},
$g:=r_{\bR_\alpha}u^+$, $h:=r_{\bR^+}(\pa_\nub u)^+$ and
$g_0\in\bW^{s-1/p}_p(\Gamma_\alpha)$, $h_0\in\bW^{s-1-1/p}_p(\Gamma_\alpha)$
are some fixed extensions of $g$ and $h$ to $\Gamma_\alpha$, then $\vf_0:=u^+ - g_0$, $\psi_0:=(\pa_t u)^+ - h_0$ are solutions to the system \eqref{e0.13}.

The system of boundary pseudodifferential equations \eqref{e0.13} has a unique pair of solutions $\varphi_0,\psi_0\in\wt{\bW}^{-1/2}(\Gamma_\alpha)
=\wt{\bH}^{-1/2}(\bR^+)$ in the classical setting $p=2$, $s=1$.
\end{theorem}

The proof of Theorem \ref{t0.5} is exposed in \S\, \ref{sect3}.

For the system \eqref{e0.13} we can remove the constraint $\dst\frac1p<s<1+\dst\frac1p$ and consider two different settings for arbitrary $r\in\bR$:
\begin{subequations}
\begin{eqnarray}\label{e0.15a}
&\vf_0\in\wt{\bW}{}^r_p(\bR^+), \quad \psi_0\in\wt{\bW}{}^{r-1}_p(\bR^+),
      \quad G\in\bW^r_p(\bR^+),\quad H \in\bW^{r-1}_p(\bR^+),\\[3mm]
\label{e0.15b}
&\vf_0\in\wt{\bH}{}^r_p(\bR^+), \quad \psi_0\in\wt{\bH}{}^{r-1}_p(\bR^+),  \quad G\in\bH^r_p(\bR^+),\quad H\in\bH^{r-1}_p(\bR^+)
\end{eqnarray}
\end{subequations}
 %
\begin{theorem}\label{t0.6}
Let $1<p<\infty$, $r\in\bR$.

The system of boundary integral equations \eqref{e0.13} is Fredholm in both the Sobolev-Slobode\v{c}kii \eqref{e0.15a} and the Bessel potential \eqref{e0.15b} space settings if and only if:
{
 \begin{eqnarray}\label{e0.16}
e^{4\pi ri}\sin^2\pi\left(\dst\frac1p-r-1\right)+
     \cos^2(\pi -\alpha)\left(\dst\frac1p-r-1\right)\not=0.
 \end{eqnarray}}
In details the condition \eqref{e0.16} is written as follows:
{
\begin{subequations}
 \begin{eqnarray}\label{e0.16a}
&&\hskip-17mm 1)\;\; r\not=\dst\frac1p-n\quad \text{and}\quad r\not=\dst\frac n4;\\
\label{e0.16b}
&&\hskip-17mm 2)\;\;\text{If}\quad r=\frac1p-n,\;\; n\not=0,\quad
     \text{then}\quad\alpha\not=\frac{2k+1}{2n}\pi,\\
\label{e0.16c}
&&\hskip-17mm 3)\;\;\text{If}\;\;r=\dst\frac n2,\;\;\text{then}\;\; p\not=\dst\frac2{n+2m+2}\;\; \text{or}\;\;\alpha\not=\dst\frac{2k+1}{2m}\pi,\\
\label{e0.16d}
&&\hskip-19mm \begin{array}{l}
4)\;\;\text{If}\quad r=\dst\frac{2n+1}4, \quad\text{then}\\
     \quad\; \alpha\not=2\pi p\dst\frac{2k-1}{4-2np-3p}\quad \text{and}
     \quad \alpha\not=2\pi\frac{2-np-2kp}{2-2np-3p},
     \end{array}
 \end{eqnarray}
\end{subequations}}
where $k,m,n=0,\pm1,\ldots.$

The system \eqref{e0.13} has the unique pair of solutions in the space settings \eqref{e0.15a} and  \eqref{e0.15b} if:
 \begin{eqnarray}
\begin{array}{l} \label{e0.17}
-\dst\frac34<r\leqslant\dst\frac1p-1 \quad\text{for}\quad
     1<p\leqslant2,\\
{\dst\frac1p-1\leqslant r<-\dst\frac14\quad\text{for}}\quad 2\leqslant p<\infty.
 \end{array}
 \end{eqnarray}
\end{theorem}

The proof of the Theorem \ref{t0.6} is exposed in \S\, \ref{sect5}.

Investigations of the boundary integral equations run into difficulties due to the absence of results on Mellin convolution equations \eqref{e0.13} in the  Sobolev-Slobode\v{c}kii \eqref{e0.15a} and the Bessel potential \eqref{e0.15b} space settings. In the present paper we apply the results on Mellin convolution equations with meromorphic kernels in the Bessel potential and Sobolev-Slobode\v{c}kii spaces  obtained recently by  R. Duduchava \cite{Du15} and V. Didenko \& R. Duduchava \cite{DD16}. We write explicitly the symbol $\cM^s_{\alpha,p}(\omega)$  of the corresponding operator as a function on the infinite rectangle $\mathfrak{R}$, {and this symbol is responsible, as usual,} for the Fredholm property and the index of the operator.

Major contribution to BVPs for elliptic equations in two and multidimensional domains with edges and cones on the boundary was made by V. Kondratjev by his celebrated paper \cite{Ko67}. The method is based on Mellin transformation and allows to find asymptotic of solutions. The approech was very popular and used intensively in the literature, see papers and monographs by P. Grisward \cite{Gr85}, M. Dauge \cite{Da88}, V. Kozlov, V. Mazya, J. Rossman \cite{KMR01}, B.W. Schulze \cite{Sc92} and many others. The investigations are mostly performed in special Kondratjev's weighted spaces, adapted to the geometry of domains with singularities.

In the recent papers \cite{CK08,CK10,CK13,CK15} L. Castro \& D. Kapanadze reduce  BVP  \eqref{e0.4}  in the $\bH^{1+\ve}(\Omega_\alpha)$ space settings to an equivalent equation with Wiener-Hopf $\pm$ Hankel operators, by manipulating  with the even and odd extensions and the reflection operators. The obtained equations were investigated in  $\bL_2(\bR^+)$ space and, in the last paper \cite{CK15}, in the special potential space, defined by the Mellin {transformations.}

In a series of papers \cite{Kru98,Kru01,Kru07,Kru09} P.A. Krutitskii investigated Boundary value problems for the Helmholtz equation in a planar 2D unbounded domain $\Omega$ outer to a finite number of finite domains and cuts with different boundary conditions. Unique solvability was proved in classical strong setting $u\in C^1(\overline{\Omega})\cap C^2(\Omega)$.

Rigorous analytical solution of the model boundary value problems with
different boundary conditions is crucial for understanding elliptic boundary value problems in Lipschitz domains (see \cite{KS03,KMR01,No58} and \cite{Me87} for the physical background and early references). In \cite{BDKT13} is described how the modern localization technique can be applied to the investigation {of BVPs in domains} with the Lipschitz boundary by reducing {them to several local} Boundary
value problems in model domains.

Model BVPs for rational angles in the classical setting are solved explicitly in \cite{ENS13a,ENS13b}. Other known results are either very limited to special situations such as the rectangular case \cite{CST04,CST06,MPST93} or rather complicated in what concerns the analytical methods \cite{KMM05,ZM00} or not describing appropriate function spaces, see, e.g., \cite{Ma59,Uf03}.  For the historical survey and for further references we recommend \cite{CK13,ZM00,Va00}.

Yet another approach, which can also be applied, is the limiting absorption principle, which is based on variational formulation and Lax-Milgram Lemma and its generalizations (see e.g., in \cite{BT01,BCC12a,BCC12b}) but, again, BVPsa are considered in the classical setting $p=2$ only.

In the 1960’s there was suggested to solve canonical diffraction problems in Sobolev spaces, based on results on pseudodifferential equations in domains with corners and, more generally, in Lipschitz domains (see papers of E. Meister \cite{Me85,Me87}, E. Meister and F.-O. Speck \cite{MS79}, W.L. Wendland \cite{WSH79}, A. Ferreira dos Santos \cite{ST89} etc.) In the book of Vasil’ev \cite{Va00} one find a considerable list of references.

There are many other papers where concrete diffraction problems are studied. We confine ourselves only with the rederence to some of them:  \cite{CK06}, \cite{CK08}, \cite{CK10} \cite{CST03}, \cite{Kru01}, \cite{Kru09}, \cite{KMM05}, \cite{Ma59}, \cite{MPST93}, \cite{MPST98}, \cite{MR96}, \cite{MSV11}, \cite{MSB11}, \cite{MSS97}, \cite{ZM00}.

\section{Boundary poseudodifferential operators}
\label{Sec1}
\setcounter{equation}{0}

Let  $\cH_k(x)$ be the fundamental solution to the Helmholtz equation
 \begin{eqnarray}\label{e1.2}
\Delta\cH_k(x) + k^2\cH_k(x)=\delta(x),\qquad x\in\bR^2,
 \end{eqnarray}
which coincides with the Hankel function of the first kind and order $0$
$\cH_k(x)=\dst\frac14H^{(1)}_0(k|x|)$. The Hankel function decays
exponentially at the infinity and has the following asymptotic
{(see \cite{GR94,Kru98}):}
\begin{eqnarray}\label{e1.4}
 H^{(1)}_0(|z|)=\dst\frac2{\pi}\ln\,|z|+{\rm const} { + \cO(|z|^2\ln\,|z|)
 \quad\hbox{\rm as}}\quad |z|\to0.
\end{eqnarray}
It is important to note, that {the asymptotic equality \eqref{e1.4} remains valid after taking any finite number of derivatives.}

Consider standard layer potential operators on the model domain $\Omega_\alpha$, the Newton, the Single and the Double layer potentials respectively
 \begin{eqnarray}\label{e1.1}
\begin{array}{l}
\N_{\Delta+k^2}\vf(x):=\dst\int_{\Omega_\alpha}
     \cH_k(x-y)\vf(y)\,dy,\\[3mm]
\V_{\Delta+k^2}\vf(x):=\dst\int_{\Gamma_\alpha}\cH_k(x-\tau)\vf(\tau)d\sigma,\\[3mm]
\W_{\Delta+k^2}\vf(x):=\dst\int_{\Gamma_\alpha}\pa_{\nub(\tau)}\cH_k(x-\tau)\vf(\tau)d\sigma,
     \qquad x\in\Omega_\alpha.
\end{array}
\end{eqnarray}

The potential operators, defined above, have standard boundedness properties in the Bessel potential spaces (see, e.g., \cite{DNS95,Du01,HW08}):
 \begin{eqnarray}\label{e1.5}
\begin{array}{rcl}
\N_{\Delta+k^2} &:&\bH_p^s(\Omega_\alpha)\longrightarrow
     \bH^{s+2}_p(\Omega_\alpha),\qquad s\in\bR,\quad 1<p<\infty,\\
\V_{\Delta+k^2}  &:&\bH_p^r(\Gm_\alpha)\longrightarrow
     \bH^{r+1+\frac1p}_p(\Omega_\alpha),\\
\W_{\Delta+k^2} &:&\bH_p^r(\Gm_\alpha)\longrightarrow
     \bH^{r+\frac1p}_p(\Omega_\alpha),\qquad \dst\frac1p-2<r<\dst\frac1p+1,\quad 1<p<\infty.
\end{array}
 \end{eqnarray}

It is well known that any solution $u\in\bH^1(\Omega_\alpha)$ to the BVP \eqref{e0.4} in the space is represented as {follows
\begin{eqnarray}\label{e2.1}
u(x)=\N_{\Delta+k^2}f(x)+\W_{\Delta+k^2}u^+(x)
     -\V_{\Delta+k^2}[\pa_\nub u]^+(x)\qquad x\in\Omega_\alpha,
 \end{eqnarray}
(see \cite{DNS95,Du01}) where densities represent} the Dirichlet $u^+$ and the Neumann $[\pa_\nub u]^+$ traces of the solution $u$ on the boundary.

Let us remind the Plemelji formulae
\begin{eqnarray}\label{e2.2}
\begin{array}{rcl}
(\W_{\Delta+k^2}\vf)^\pm(t)=\pm\dst\frac12\vf(t)+\W_{{\Delta+k^2},0}\vf(t),\\[3mm]
     (\pa_\nub\V_{\Delta+k^2}\psi)^\pm(t)=\mp\dst\frac12
     \psi(t)+\W^*_{{\Delta+k^2},0}\psi(t),\\[3mm]
(\pa_\nub\W_{\Delta+k^2}\psi)^\pm(t)=\V_{{\Delta+k^2},+1}\psi(t),\\[3mm]
(\V_{\Delta+k^2}\vf)^\pm(t)=\V_{{\Delta+k^2},-1}\vf(t) \qquad t\in\Gamma_\alpha:=\pa\Omega_\alpha,
 \end{array}
 \end{eqnarray}
where the pseudodifferential operators  ($\Psi$DO)
 \begin{eqnarray}\label{e2.3}
\begin{array}{l}
\V_{\Delta+k^2,-1}\vf(t):=\dst\int_{\Gamma_\alpha}\cH_k(t-\tau)\vf(\tau)d\sigma,\\[3mm]
\W_{\Delta+k^2,0}\vf(t):=\dst\int_{\Gamma_\alpha}\pa_{\nub(\tau)}\cH_k(t-\tau)\vf(\tau)d\sigma,\\[3mm]
\W^*_{\Delta+k^2,0}\vf(t):=\dst\int_{\Gamma_\alpha}\pa_{\nub(t)}\cH_k(t-\tau)\vf(\tau)d\sigma,\\[3mm]
\V_{\Delta+k^2,+1}\vf(t):=\dst\int_{\Gamma_\alpha}\pa_{\nub(t)}\pa_{\nub(\tau)}\cH_k(t-\tau)
     \vf(\tau)d\sigma, \qquad t\in\Gamma_\alpha
\end{array}
\end{eqnarray}
of orders $-1$, $0$, $0$ and $+1$, are associated with the layer potentials of the Helmholtz equation.  Due to the asymptotic \eqref{e1.4}, the operator $\V_{\Delta+k^2,-1}$ has weakly singular kernel and the integral exists in the Lebesgue sense, while the operators $\W_{\Delta+k^2,0}$ and $\W^*_{\Delta+k^2,0}$ have singular kernel of order $-1$ and the integrals exists in the Cauchy Mean Value sense.

To explain in which sense is understood the hypersingular integral operator $\V_{\Delta+k^2,+1}\vf(t)$, let us recall the following equality
\begin{eqnarray}\label{e2.4a}
\Delta+k^2=\partial^2_1+\partial^2_2+k^2=\partial^2_\nub+\partial^2_\ell+k^2,
 \end{eqnarray}
{
where $\nub(t)$ is the unit normal vector field, $\partial_\nub$ is the normal derivative (see \eqref{e0.1}--\eqref{e0.3}),

From \eqref{e1.2} and \eqref{e2.4a} follows the equality
 \[
\delta=(\Delta+k^2)\cH_k=\partial^2_\nub\cH_k +(\partial^2_\ell+k^2)\cH_k,
 \]
which we use to prove the following:
\begin{eqnarray}\label{e2.4c}
\partial_{\nub(x)}\partial_{\nub(y)}\cH_k(x-y)=-\partial^2_{\nub(y)}\cH_k(x-y)
     =-\delta(x-y)+(\partial^2_{\ell(y)}+k^2)\cH_k(x-y).
 \end{eqnarray}

Due to the equality \eqref{e2.4c} and the integration by parts formula for the tangential differential operator (see \cite{Du01,DMM06})
\begin{eqnarray}\label{e1.10}
\dst\int_{\Gamma_\alpha}\pa_{\ell(\tau)}\psi(\tau)\vf(\tau)d\sigma          =-\dst\int_{\Gamma_\alpha}\psi(\tau)\pa_{\ell(\tau)}\vf(\tau)d\sigma.
 \end{eqnarray}
the hypersingular operator $\V_{\Delta+k^2,+1}$ is represented as
 \begin{eqnarray}\label{e2.5}
\V_{\Delta+k^2,+1}\vf(t)&\hskip-3mm:=&\hskip-3mm\dst\int_{\Gamma_\alpha}
     \pa_{\nub(t)}\pa_{\nub(\tau)}\cH_k(t-\tau)\vf(\tau)d\sigma\nonumber\\
&\hskip-3mm=&\hskip-3mm-\vf(t)+\dst\int_{\Gamma_\alpha}(\pa^2_{\ell(\tau)}
     +k^2)\cH_k(t-\tau)\vf(\tau)d\sigma\nonumber\\
&\hskip-3mm=&\hskip-3mm-\vf(t)-\dst\int_{\Gamma_\alpha}\pa_{\ell(\tau)}
     \cH_k(t-\tau)\pa_{\ell(\tau)}\vf(\tau)d\sigma \nonumber\\
&&+k^2\dst\int_{\Gamma_\alpha}\cH_k(t-\tau)\vf(\tau)d\sigma, \quad
     t\in\Gamma_\alpha.
\end{eqnarray}
Accoding to the obtained equality \eqref{e2.5} the operator $\V_{\Delta+k^2,+1}$ is a sum of singular integral operator applied to the tangential derivative $\pa_\ell\vf$ of the density and the regular integral applied to $\vf$ itself.

The following pseudodifferential operators
 \begin{eqnarray}\label{e2.6}
\begin{array}{l}
\V_{\Delta,-1}\vf(t):=\dst\frac1{2\pi}\int_{\Gamma_\alpha}\ln|t-\tau|\vf(\tau)d\sigma,\\[3mm]
\W_{\Delta,0}\vf(t):=\dst\frac1{2\pi}\dst\int_{\Gamma_\alpha}\pa_{\nub(\tau)}\ln|t-\tau|\vf(\tau)d\sigma,\\[3mm]
\W^*_{\Delta,0}\vf(t):=\dst\frac1{2\pi}\dst\int_{\Gamma_\alpha}\pa_{\nub(t)}\ln|t-\tau|\vf(\tau)d\sigma,\\[3mm]
\V_{\Delta,+1}\vf(t):=\dst\frac1{2\pi}\dst\int_{\Gamma_\alpha}\pa_{\nub(t)}\pa_{\nub(\tau)}\ln|t-\tau|
     \vf(\tau)d\sigma, \qquad t\in\Gamma_\alpha
\end{array}
\end{eqnarray}
of orders $-1$, $0$, $0$ and $+1$, are associated with the Laplace equation (see \cite{DNS95,Du01,HW08})
\begin{eqnarray*}
    \Delta u(x) = 0, \qquad x\in\bR^2.
 \end{eqnarray*}
which has the logarithmic fundamental solution
\begin{eqnarray*}
  \cK_\Delta(x):=\frac1{2\pi}\ln|x|,\qquad   \Delta\cK_\Delta(x) = \delta(x), \qquad x\in\bR^2,
 \end{eqnarray*}

The pseudodifferential operators defined above have standard mapping properties (see \cite{DNS95,Du01,HW08}):
 \begin{eqnarray}\label{e2.7}
\begin{array}{rcl}
\V_{\Delta+k^2,-1},\ \V_{\Delta,-1} &:&\bH_p^s(\Gm_\alpha)\longrightarrow
     \bH^{s+1}_p(\Gm_\alpha),\\[3mm]
\W_{\Delta+k^2,0},\ \W_{\Delta,0} &:&\bH_p^s(\Gm_\alpha)\longrightarrow
     \bH^s_p(\Gm_\alpha),\\[3mm]
\W^*_{\Delta+k^2,0},\ \W^*_{\Delta,0} &:&\bH_p^s(\Gm_\alpha)\longrightarrow
     \bH^s_p(\Gm_\alpha),\\[3mm]
\V_{\Delta+k^2,+1},\ \V_{\Delta,+1} &:&\bH_p^s(\Gm_\alpha)\longrightarrow
     \bH^{s-1}_p(\Gm_\alpha)
\end{array}
 \end{eqnarray}
for $1<p<\infty$, $\dst\frac1p-2<r<\dst\frac1p+1$.
 %
\begin{lemma}[Lemma 1.1, \cite{DD16}]\label{l1.1}
Let $1<p<\infty$, $\dst\frac1p-2<r<\dst\frac1p+1$ and either $\bR_0=\bR^+$, $\bR_1=\bR_\alpha$ or vice versa $\bR_0=\bR_\alpha$, $\bR_1=\bR^+$. Let, respectively, $r_j\;:\;\bH_p^s(\Gamma_\alpha)\to\bH_p^s(\bR_j)$, $j=0,1$, be the corresponding restriction operators. Then the differences
 \begin{eqnarray}\label{e2.7a}
\begin{array}{rcl}
\T_1:=r_1[\V_{\Delta+k^2,-1} - \V_{\Delta,-1}]r_0 &:&\wt{\bH}_p^s(\bR_0)\longrightarrow
     \bH^{s+1}_p(\bR_1),\\[3mm]
\T_2:=\W_{\Delta+k^2,0} -  \W_{\Delta,0} &:&\bH_p^s(\Gm_\alpha)\longrightarrow
     \bH^s_p(\Gm_\alpha),\\[3mm]
\T_3=\T_2^*:=\W^*_{\Delta+k^2,0} - \W^*_{\Delta,0} &:&\bH_p^s(\Gm_\alpha)\longrightarrow
     \bH^s_p(\Gm_\alpha),\\[3mm]
\T_4:=r_1[\V_{\Delta+k^2,+1} -  \V_{\Delta,+1}]r_0 &:&\wt{\bH}_p^{s+1}(\bR_0)\longrightarrow
     \bH^s_p(\bR_1)
\end{array}
 \end{eqnarray}
are locally compact operators: the operators $v\T_j$, $j=1,2,3,4$, are compact for arbitrary function $v\in C^\infty_0(\Gamma_\alpha)$ with a compact support.
\end{lemma}

Next we will write {some pseudodifferential} operators (PsDOs) in \eqref{e2.6} in explicit form for the later use in \S\, 3.. First let us   consider the PsDOs $r_{\bR^+}\V_{\Delta,+1}r_{\bR_\alpha}$. By applying the equality
 \[
\partial_{\nub(x)}\partial_{\nub(y)}\ln|x-y|=-\partial^2_{\nub(y)}\ln|x-y|
     =-\delta(x-y)+\partial^2_{\ell(y)}\ln|x-y|,
 \]
proved similarly to \eqref{e2.5},  we get:
 \begin{eqnarray*}
&&\V_{\Delta,+1}\vf(t):=\dst\int_{\Gamma_\alpha}\pa_{\nub(t)}\pa_{\nub(\tau)}
     \ln|t-\tau|\vf(\tau)d\sigma=-\vf(t)+\dst\int_{\Gamma_\alpha}\pa^2_{\ell
     (\tau)}\ln|t-\tau|\vf(\tau)d\sigma\nonumber\\
&&\hskip30mm=-\vf(t)-\dst\int_{\Gamma_\alpha}\pa_{\ell(\tau)}\ln|t-\tau|
     \pa_{\ell(\tau)}\vf(\tau)d\sigma, \quad t\in\Gamma_\alpha.
\end{eqnarray*}

By using the parametrization  $x=(x_1,x_2)^\top=(t,0)^\top$ of {$\bR^+$, the} parametrization  $y=(y_1,y_2)^\top=(\tau\cos\,\alpha,\tau\sin\,\alpha)^\top$ of $\bR_\alpha$, recalling that $\bR_\alpha$ is oriented from $-\infty$ to $0$ and using the equality \eqref{e0.3}, equalities
 \begin{eqnarray}\label{e1.16}
{\pa_{\ell(y)}}=-\cos\,\alpha\,\pa_{y_1}-\sin\,\alpha\,\pa_{y_2},\quad
     \ln|x-y|=\frac12\ln\big[(x_1-y_1)^2 + (x_2-y_2)^2\big]
  \end{eqnarray}
for $t\in\bR^+$, $y\in\Gamma_\alpha$, we proceed as follows:
\begin{align*}
\pa_{\nub(t)} & \pa_{\nub(y)}\cK_\Delta(t\!-\!y)=\pa_{\nub(x)}
     \pa_{\nub(y)}\cK_\Delta(x\!-\!y)\Big|_{\begin{array}{l}\scriptstyle x=(t,0)\\[-2mm]
     \scriptstyle y=(\tau\cos\,\alpha,\tau\sin\,\alpha)\end{array}}\\
&=-\!\pa_{x_2}(-\sin\,\alpha\,\pa_{y_1}+\cos\,\alpha\,\pa_{y_2})
     \cK_\Delta(x\!-\!y)\Big|_{\begin{array}{l}\scriptstyle x=(t,0)\\[-2mm]
     \scriptstyle y=(\tau\cos\,\alpha,\tau\sin\,\alpha)\end{array}}\\
&=\left\{-\sin\,\alpha\,\pa_{y_1}\pa_{y_2}+\cos\,\alpha\,
     \pa_{y_2}^2\right\}\cK_\Delta(x\!-\!y)\Big|_{\begin{array}{l}
     \scriptstyle x=(t,0)\\[-2mm]
     \scriptstyle y=(\tau\cos\,\alpha,\tau\sin\,\alpha)\end{array}}\\
&=\left[\cos\,\alpha\Delta\cK_\Delta(x-y)-\pa_{y_1}\left\{\cos\,\alpha\,
     \pa_{y_1}+\sin\,\alpha\,\pa_{y_2}\right\}\cK_\Delta(x-y)
     \right]\Big|_{\begin{array}{l}\scriptstyle x=(t,0)\\[-2mm]
     \scriptstyle y=(\tau\cos\,\alpha,\tau\sin\,\alpha)\end{array}}\\
&=\left[\cos\,\alpha\delta(x-y)+\pa_{y_1}\pa_{\ell(y)}\cK_\Delta(x-y)\right]
     \Big|_{\begin{array}{l}\scriptstyle x=(t,0)\\[-2mm]
     \scriptstyle y=(\tau\cos\,\alpha,\tau\sin\,\alpha)\end{array}}\\
&=\left[\cos\,\alpha\,\delta(0)+\frac1{4\pi}\pa_{\ell(y)}\pa_{y_1}
     \ln\big[(x_1-y_1)^2 + (x_2-y_2)^2\big]\right]
     \Big|_{\begin{array}{l}\scriptstyle x=(t,0)\\[-2mm]
     \scriptstyle y=(\tau\cos\,\alpha,\tau\sin\,\alpha)\end{array}}\\
&=\left[\cos\,\alpha\,\delta(0)-\frac1{2\pi}\pa_{\ell(y)}\frac{x_1-y_1}
     {2\pi[(x_1-y_1)^2 + (x_2-y_2)^2]}\right]
     \Big|_{\begin{array}{l}\scriptstyle x=(t,0)\\[-2mm]
     \scriptstyle y=(\tau\cos\,\alpha,\tau\sin\,\alpha)\end{array}}
\end{align*}
Now integrating by parts (see \eqref{e1.10}) we continue as follows:
\begin{align}\label{e1.17}
r_{\bR^+}\V_{\Delta,+1}r_{\bR_\alpha}v(t)&=\frac1{2\pi}
     r_{\bR^+}\dst\int_{\bR_\alpha}\frac{(x_1-y_1)\pa_{\ell(y)}v(y)d\sigma}
     {(x_1-y_1)^2 + (x_2-y_2)^2} \Big|_{\begin{array}{l}\scriptstyle x=(t,0)\\[-2mm]
     \scriptstyle y=(\tau\cos\,\alpha,\tau\sin\,\alpha)\end{array}} \hskip-10mm\nonumber\\
&=-\frac1{2\pi}\dst\int_0^\infty\frac{t-\tau\cos\,\alpha}
     {t^2 + \tau^2 - 2t\tau\,\cos\alpha}(\J_\alpha\pa_\ell v)(\tau)d\tau
     \nonumber\\
&=-\frac1{4\pi}\int_0^\infty\left[\dst\frac1{
     t-e^{i\alpha}\tau}+\frac1{t-e^{-i\alpha}\tau}\right]
     (\J_\alpha\pa_\ell v)(\tau)d\tau,\nonumber\\
&=\frac14\left[\K_{e^{i\alpha}}
     +\K_{e^{-i\alpha}}\right]\pa_\tau v_1(t),\qquad t\in\bR^+,
\end{align}
since $(\J_\alpha\pa_\ell v)(\tau)=-(\pa_\tau v_1)(\tau)$, where $v_1:=\J_\alpha v$.

The formula
 \[
\J_\alpha r_{\bR_\alpha}\V_{\Delta,+1}r_{\bR^+}w(t)
     =-\frac14\left[\K_{e^{i\alpha}}+\K_{e^{-i\alpha}}
     \right](\pa_\tau w)(t),  \quad t\in\bR^+
 \]
is proved similarly.

Now we look to the singular integral operators $r_{\bR^+}\pa_\ell\V_{\Delta,-1}r_{\bR_\alpha}$ and $r_{\bR_\alpha}\pa_\ell\V_{\Delta,-1}r_{\bR^+}$. We proceed as in \eqref{e1.17}:
 \begin{eqnarray}\label{e1.19}
\J_\alpha r_{\bR_\alpha}\pa_\ell\V_{\Delta,-1}r_{\bR^+}w(t)&\hskip-3mm=&\hskip-3mm\frac1{2\pi}\J_\alpha
     r_{\bR_\alpha}\dst\int_{\bR^+}\pa_{\ell(x)}\ln|x-y|\,w(y)d\sigma\nonumber\\
&\hskip-3mm=&\hskip-3mm-\frac1{2\pi}\J_\alpha r_{\bR_\alpha}\dst\int_{\bR^+}\frac{\cos\,\alpha(x_1 - \tau)
     +x_2\sin\,\alpha}{(x_1-\tau)^2 + x^2_2}w(\tau)\,d\tau\nonumber\\
&\hskip-3mm=&\hskip-3mm-\frac1{2\pi}\dst\int_0^\infty\frac{\cos\,\alpha(t\cos\,\alpha-\tau) + t\,\sin^2\alpha}
      {(t\,\cos\alpha-\tau)^2 + t\sin^2\alpha}w(\tau)d\tau\nonumber\\
&\hskip-3mm=&\hskip-3mm-\frac1{2\pi}\dst\int_0^\infty\frac{t-\tau\,\cos\alpha}
      {(t\,\cos\alpha - \tau)^2 + t^2\sin^2\alpha}w(\tau)d\tau\nonumber\\
&\hskip-3mm=&\hskip-3mm-\frac1{4\pi}\int_0^\infty\left[\dst\frac1{t-e^{i\alpha}\tau}
     +\frac1{t-e^{-i\alpha}\tau}\right]w(\tau)d\tau,\nonumber\\
&\hskip-3mm=&\hskip-3mm-\frac14\left[\K_{e^{i\alpha}}+\K_{e^{-i\alpha}}\right]w(t),
      \quad t\in\bR^+.
\end{eqnarray}

The formulae
 \begin{eqnarray}\label{e2.12a}
&&\hskip-20mm r_{\bR^+}\pa_\ell\V_{\Delta,-1}r_{\bR_\alpha}w(t)
     =\frac14\left[\K_{e^{i\alpha}}+\K_{e^{-i\alpha}}
     \right]\J_\alpha w(\tau),   \quad t\in\bR^+
\end{eqnarray}
is proved similarly.

For the singular integral operator $\W_{\Delta,0}$ we proved the following:
 \begin{eqnarray}\label{e2.8a}
&&\hskip-20mm r_{\bR^+}\W_{\Delta,0}r_{\bR_\alpha}\vf(t)
     =- \J_\alpha r_{\bR_\alpha}\W_{\Delta,0}r_{\bR^+}\vf(t)
     =\frac1{4i}\left[e^{i\alpha}\K_{e^{i\alpha}}-e^{-i\alpha}
     \K_{e^{-i\alpha}}\right]\vf_1(t),\\
\label{e2.8b}
&&\hskip-20mm r_{\bR^+}\W_{\Delta+k^2}r_{\bR^+}=r_{\bR^+}\W_{\Delta}r_{\bR^+}
     =r_{\bR_\alpha}\W_{\Delta+k^2}r_{\bR_\alpha}
     =r_{\bR_\alpha}\W_{\Delta}r_{\bR_\alpha}=0,\\
&&\hskip60mm \vf_1(t):=(\J_\alpha\vf)(t),  \qquad t\in\bR^+, \nonumber
\end{eqnarray}
where  $\J_\alpha$ is the pull back operator (see \eqref{e0.8}) and
$r_{\bR^+}$ and $r_{\bR_\alpha}$ are the restriction operators to the
spaces on the corresponding subsets $\bR^+$ and $\bR_\alpha$. {We drop the proofs of \eqref{e2.8a} and \eqref{e2.8b} because these formulae are not applied in the present manuscript.}

For the dual operator $\W^*_{\Delta,0}$ we get:
 \begin{eqnarray}\label{e2.9a}
&&\hskip-15mm r_{\bR_\alpha}\W^*_{\Delta,0}r_{\bR^+}\vf(t)=- \J_\alpha
     r_{\bR^+}\W^*_{\Delta,0}r_{\bR^+}\vf(t)
     =\frac1{4i}\left[\K_{e^{i\alpha}}-\K_{e^{-i\alpha}}\right]\vf(t),
     \quad t\in\bR^+,\\
\label{e2.9b}
&&\hskip-15mm r_{\bR^+}\W^*_{\Delta+k^2}r_{\bR^+}=r_{\bR^+}
     \W^*_{\Delta}r_{\bR^+}=r_{\bR_\alpha}\W^*_{\Delta+k^2}r_{\bR_\alpha}
     =r_{\bR_\alpha}\W^*_{\Delta}r_{\bR_\alpha}=0.
\end{eqnarray}

\section{Boundary integral equation of the model problem}
\label{sect3}
\setcounter{equation}{0}

{\bf Proof of Theorem \ref{t0.5}:} The boundary data  $g\in\bW^{s-1/p}_p(\bR_\alpha)$ and $h\in\bW^{s-1-1/p}_p(\bR^+)$ of the BVP \eqref{e0.4} in the non-classical formulation \eqref{e0.6} are defined initially on the parts of the boundary $\bR_\alpha$ and $\bR^+$, respectively.  Let $g_0\in\bW^{s-1/p}_p(\Gamma_\alpha)$ and $h_0\in\bW^{s-1-1/p}_p(\Gamma_\alpha)$ be some fixed extensions of these boundary data to the entire boundary $\Gamma_\alpha=\bR^+\cup\bR_\alpha$. We remind, that the spaces $\wt{\bW}^{s-1/p}_p(\bR_\alpha)$ and $\wt{\bW}^{s-1/p}_p(\bR^+)$ are subsets of $\bW^{s-1/p}_p(\Gamma_\alpha)$ and functions from $\wt{\bW}^s_p(\bR^+)$ and $\wt{\bW}^s_p(\bR_\alpha)$ are extended by $0$ to  $\bR_\alpha$ and to $\bR^+$, respectively. The difference between of two such extensions belong to the spaces $\wt{\bW}^{s-1/p}_p(\bR_\alpha)$ and $\wt{\bW}^{s-1-1/p}_p(\bR^+)$ respectively. Therefore, we should look for two unknown functions $\varphi_0\in\wt{\bW}^{s-1/p}_p(\bR^+)$ and $\psi_0\in\wt{\bW}^{s-1-1/p}_p(\bR_\alpha)$, such that for $g_0+\varphi_0$ and $h_0+\psi_0$ the boundary conditions in \eqref{e0.4} hold on the entire boundary. i.e., for any solution $u(x)$ to the BVP \eqref{e0.4}, there holds 
\begin{eqnarray}\label{e3.1}
\begin{array}{c}
u^+(t)=g_0(t)+\varphi_0(t)=\left\{\begin{array}{ll} g_0(t)+\varphi(t)\quad &{\rm if}\quad
     t\in\bR^+,\\[3mm]
g_0(t)\quad &{\rm if}\quad t\in\bR_\alpha,
\end{array}\right.\\ \\
(\pa_\nub u)^+(t)=h_0(t)+\psi_0(t)=\left\{\begin{array}{ll} h_0(t)
     \quad &{\rm if}\quad t\in\bR^+,\\[3mm]
h(t)+\psi_0(t)\quad &{\rm if}\quad t\in\bR_\alpha.
\end{array}\right.
\end{array}
\end{eqnarray}

By introducing the boundary values of a solution \eqref{e3.1} of the BVP \eqref{e0.4} into the representation formula \eqref{e2.1} we get the
following representation of a solution:
\begin{eqnarray}\label{e3.2}
u(x)=\N_{\Delta+k^2} f(x)+\W_{\Delta+k^2}[g_0+\varphi_0](x)-\V_{\Delta+k^2}[h_0+\psi_0](x),\qquad x\in\cC.
\end{eqnarray}
The known and unknown functions in \eqref{e3.1} and \eqref{e3.2} belong to the following spaces
\begin{eqnarray}\label{e3.3}
g_0\in\bW^{s-1/p}_p(\Gamma_\alpha), \quad h_0\in\bW^{s-1-1/p}_p(\Gamma_\alpha), \quad
\varphi_0\in\wt{\bW}^{s-1/p}_p(\bR^+), \quad \psi_0\in\wt{\bW}^{s-1-1/p}_p(\bR_\alpha).
\end{eqnarray}

By applying the boundary conditions from \eqref{e0.4} to \eqref{e3.2} and the Plemelji formulae \eqref{e2.2} we get the following:
\[
\begin{array}{r}
\left\{\begin{array}{l}
g_0(t)+\varphi_0(t)=u^+(t)={(\N_{\Delta+k^2} f)^+}+\dst\frac12(g_0(t)+\varphi_0(t))\\
\hskip20mm+\W_{\Delta+k^2,0}[g_0+\varphi_0](t)-\V_{\Delta+k^2,-1}[h_0+\psi_0](t),\\[2mm]
h_0(t)+\psi_0(t)=(\pa_\nub u)^+(t)={(\pa_\nub \N_{\Delta+k^2} f)^+}+\V_{\Delta+k^2,+1}[g_0+\varphi_0](t)\\
\hskip20mm+\dst\frac12(h_0(t) +\psi_0(t))-\W^*_{\Delta+k^2,0}[h_0+\psi_0](t),\qquad t\in\Gamma.
\end{array}\right.
\end{array}
 \]
Rearranging the known and unknown functions the system acquires the following form
\begin{eqnarray}\label{e3.4}
\left\{\begin{array}{ll}\dst\frac12\varphi_0 - \W_{\Delta+k^2,0}r_{\bR^+}
      \varphi_0 + \V_{\Delta+k^2,-1}r_{\bR_\alpha}\psi_0=G_0,\\[3mm]
\dst\frac12\psi_0 + \W^*_{\Delta+k^2,0}r_{\bR_\alpha}\psi_0
     -\V_{\Delta+k^2,+1}r_{\bR^+}\varphi_0=H_0
     \qquad&\text{on}\quad\Gamma_\alpha=\partial\Omega_\alpha, \end{array}\right.
\end{eqnarray}
where $G_0$ and $H_0$ are exposed in \eqref{e0.13} and we used the properties $r_{\bR^+}\varphi_0=\varphi_0$, $r_{\bR_\alpha}\psi_0=\psi_0$.

By applying the restriction $r_{\bR^+}$ to the both parts of the first equation in \eqref{e3.4} and the  restriction $r_{\bR_\alpha}$ to the second one and by recalling the equalities $r_{\bR^+}\W_{\Delta+k^2}r_{\bR^+}
=r_{\bR_\alpha}\W^*_{\Delta+k^2}r_{\bR_\alpha}=0$ (cf. \eqref{e2.8a}, \eqref{e2.9a}) we arrive to the system \eqref{e0.13}.

There is the full equivalence between the solvability of the system \eqref{e3.4} and the sol\-vability of the BVP \eqref{e0.4}, given by the representation formula \eqref{e3.2}. Therefore the unique solvability of the BVP \eqref{e0.4} implies the unique solvability of the system \eqref{e3.4} and vice versa, The concluding assertion of Theorem \ref{t0.5} follows then from Theorem \ref{t0.1}.      \QED

In the formulation and the proof of the next Lemma \ref{l3.1} we use
localization and quasi-localization principle. A quasi-localization means "freezing coefficients" and changing underling contours and surfaces by an isomorphic but simpler ones. For details of a quasi-localization we refer the reader to the papers \cite{Si65}  and  \cite{CDS03}, where the quasi-localization is well described for singular integral operators and for BVPs, respectively. We also refer to \cite[\S\, 3]{Du15}, where is exposed a short introduction to quasi-localization.

Let us agree to understand under local equivalence and local quasi-equivalence of equations the local equivalence of the corresponding operators in the corresponding spaces.                  %
 %
 \begin{lemma}\label{l3.1}
System of the pseudodifferential equation \eqref{e0.13} and the system
\begin{eqnarray}\label{e3.6}
\left\{\begin{array}{ll}\dst\frac12\varphi_+(t) + r_{\bR^+}\V_{\Delta,-1}
     r_{\bR_\alpha}\psi_-(t)=G_1(t),\qquad &  t\in\bR^+,\\[3mm]
\dst\frac12\psi_-(t) - r_{\bR_\alpha}\V_{\Delta,+1}r_{\bR^+}\varphi_+(t)
     =H_1(t),\qquad & t\in\bR_\alpha, \end{array}\right. \\[3mm]
\varphi_+\in\wt{\bW}^{s-1/p}_p(\bR^+), \qquad
\psi_-\in\wt{\bW}^{s-1-1/p}_p(\bR_\alpha),\nonumber\\[2mm]
G_1\in\bW^{s-1/p}_p(\bR^+), \qquad
H_1\in\bW^{s-1-1/p}_p(\bR_\alpha).\nonumber
 \end{eqnarray}
are  locally equivalent at $0$.

At any other point $x\in\Gamma_\alpha\setminus{0}\cup\{+\infty,e^{i\alpha}\infty\}$, including the both infinity points $+\infty$ and $x=e^{i\alpha}\infty:=\lim_{y\to+\infty}
e^{-i\alpha}y$, the system \eqref{e0.13}
is locally quasi-equivalent to the trivial system
\begin{eqnarray}\label{e3.7}
\begin{array}{l}
\dst\frac12\varphi=H_2,  \quad \varphi,H_2\in\bW^{s-1/p}_p(\bR)
     \qquad \text{for}\quad x\in\bR^+,\\[4mm]
\dst\frac12\psi=G_2,  \quad \psi,G_2\in\bW^{s-1/p-1}_p(\bR)\qquad
     \text{for}\quad x\in\bR_\alpha.
\end{array}
\end{eqnarray}
\end{lemma}
{\bf Proof:} The systems \eqref{e0.13} and \eqref{e3.6} are locally equivalent at $0$, because the differences
 \[
\begin{array}{rcl}
\T_1:=r_{\bR^+}[\V_{\Delta+k^2,-1} - \V_{\Delta,-1}]r_{\bR_\alpha}&:&
     \wt{\bW}_p^r(\bR_\alpha)\longrightarrow\bW^{r+1}_p(\bR^+),\\[3mm]
\T_4:=r_{\bR_\alpha}[\V_{\Delta+k^2,+1} - \V_{\Delta,+1}]r_{\bR^+}
     &:&\wt{\bW}_p^{r+1}(\bR^+)\longrightarrow\bW^r_p(\bR_\alpha)
\end{array}
\]
are locally compact for all $r\in\bR$ due to Lemma \ref{l1.1} and compact operators are locally equivalent to $0$.

Now let us describe the local quasi-equivalent systems of \eqref{e0.13} at $x\in\bR^+\cup\{+\infty\}$. Operators $\A_1:=\dst\frac12r_{\bR_\alpha}$
$\A_2:=r_{\bR^+}\V_{\Delta+k^2,+1}r_{\bR_\alpha}$, $\A_3:=r_{\bR_\alpha}\V_{\Delta+k^2,+1}r_{\bR^+}I$ and are locally
quasi-equivalent to $0$ since $v_x\A_1=\A_1v_xI=v_x\A_3=\A_2v_xI=0$ while the operators $v_x\A_2$ and $\A_3v_xI$ are compact for all $v_x\in C^\infty(\bR)$, $O\not\in\supp\,v_x$. Compact operator are, as mentioned already, locally quasi-equivalent to $0$. The identity operator $\dst\frac12r_{\bR^+}$ is locally quasi equivalent to the identity $\dst\frac12I$ in the space on the entire axes $\bR$.

Thus, the local quasi-equivalence of the system \eqref{e0.13} and the first equation in \eqref{e3.7} at $x\in\bR^+$ follows.

The local quasi-equivalence of the system \eqref{e0.13} and the second
equation in \eqref{e3.7} at $x\in\bR_\alpha$ is proved similarly, by
using the pull back operator $\J_\alpha$ (see \eqref{e0.8}).         \QED
 %
 \begin{lemma}\label{l3.2}
The system of pseudodifferential equation \eqref{e0.13} is Fredholm if
and only if the system of pseudodifferential equation
\begin{eqnarray}\label{e3.10}
&&\left\{\begin{array}{ll}
\vf(t)-\dst\frac12\left[\K_{e^{i\alpha}}+\K_{e^{-i\alpha}}\right]\psi(\tau)
     d\tau=G(t),\\[3mm]
\psi(t)+\dst\frac12\left[\K_{e^{i\alpha}}+\K_{e^{-i
     \alpha}}\right]\vf(\tau)d\tau=H(t), \qquad t\in\bR^+,
\end{array}\right.\\[2mm]
&& \varphi,\; \psi\in\wt{\bW}^{s-1-1/p}_p(\bR^+),\qquad G,\;
H\in\bW^{s-1-1/p}_p(\bR^+) \nonumber
\end{eqnarray}
is locally invertible at $0$, where
\begin{equation}\label{e3.11}
\mathbf{K}^1_c\phi(t):=\displaystyle\frac1\pi\int\limits_0^\infty
     \frac{\phi(\tau)\,d\tau}{t-c\,\tau}, \qquad 0<|\arg\,c|<2\pi, \quad\phi\in\bL_p(\bR^+).
\end{equation}
is the Mellin convolutions operator (see \cite{Du79,Du84b,Du86,Du82}).
\end{lemma}
{\bf Proof:} Due to the main principle of the quasi-localization
(see Proposition 3.4 in \cite{Du15}) the system \eqref{e0.13}
is Fredholm if and only if locally quasi-equivalent systems
(equations) is locally invertible at each point of the compactification of $\Gamma_\alpha$ which includes the infinite points, i.e., for each
$x\in\Gamma_\alpha\cup\{+\infty\}\cup\{e^{\i\alpha}\infty\}$.

The systems \eqref{e3.7} is obviously uniquely solvable (the corresponding operators are invertible).

Thus, the system \eqref{e0.13} is Fredholm if and only if the system
\eqref{e3.6} is locally invertible at $0$.

Equivalence of the local solvability of the systems \eqref{e3.6} and
\eqref{e3.10} is proved as follows.

Multiply both equations in \eqref{e3.6} by 2, apply to the first equation the differentiation $\pa_t$, replace $\vf:=\pa_t\vf_0$,  apply to the second equation the  $\J_\alpha$ (see \eqref{e0.8}) and replace $\psi=\J_\alpha\psi_0$, also under the integral. Now the system \eqref{e3.10} is derived easily from \eqref{e3.6} with the help of formulae \eqref{e1.17} and \eqref{e1.19}.

To prove the local equivalence at $0$ of the systems  \eqref{e3.6} and
\eqref{e3.10} note, that the multiplication by $2$ and the pull back
operator $\J_\alpha$ are invertible. As for the differentiation
 \[
\pa_t:=\dst\frac d{dt}\;:\;\bW^r_p(\bR^+)\to\bW^{r-1}_p(\bR^+),\qquad
\pa_t\;:\;\wt\bW^r_p(\bR^+)\to \wt\bW^{r-1}_p(\bR^+)
 \]
it is locally invertible at any finite point $x\in\bR$ because the operators
 \[
\pa_t-iI\;:\;\bW^r_p(\bR^+)\to\bW^{r-1}_p(\bR^+),\qquad
\pa_t+iI\;:\;\wt\bW^r_p(\bR^+)\to \wt\bW^{r-1}_p(\bR^+)
 \]
are isomorphisms (represent the Bessel potentials; see \cite[Lemma 5.1]{Du79}). On the other hand, the embeddings
 \[
iI\;:\;\bW^r_p(\bR^+)\to\bW^{r-1}_p(\bR^+),\qquad
iI\;:\;\wt\bW^r_p(\bR^+)\to \wt\bW^{r-1}_p(\bR^+)
 \]
are locally compact due to the Sobolev's embedding theorem and the
compact perturbation can not influence the local invertibility.   \QED

\section{Mellin convolution operators in the Bessel potential spaces}
\label{sect4}
\setcounter{equation}{0}

The results of the foregoing two sections together with the results on a Banach algebra generated by Mellin and Fourier convolution operators (see \cite{Du87}) allow the investigation of the Fredholm properties of lifted Mellin convolution operators. For this we write the symbol of a model operator
\begin{equation}\label{e4.1}
\mathbf{A}:=d_0I + \sum_{j=1}^nd_j\mathbf{K}^1_{c_j},\qquad 0<\arg\,c_j<2\pi, \quad d_0,d_j\in\bC,\quad j=1,\ldots,n,
\end{equation}
in the Bessel potential spaces setting
$\A\;:\;\mathbb{H}^s_p(\mathbb{R}^+)\to \mathbb{H}^s_p( \mathbb{R}^+)$ which is compiled of the identity $I$ and Mellin convolution operators $\mathbf{K}^1_{c_1},\ldots,\mathbf{K}^1_{c_n}$ with meromorphic kernels.

To expose the symbol of the operator \eqref{e4.1}, consider the infinite
clockwise oriented ``rectangle'' $\mathfrak{R}:=\Gamma_1\cup\Gamma_2^-\cup\Gamma_2^+ \cup\Gamma_3$, where (cf. Figure 2)
\[  \Gamma_1:=\{\infty\}\times\overline{\mathbb{R}},\qquad\Gamma^\pm_2
     :=\overline{\mathbb{R}}^+\times\{\pm\infty\},\qquad
            \Gamma_3:=\{0\}\times\overline{\mathbb{R}}.
\]
\setlength{\unitlength}{0.4mm}
\vskip7mm
\hskip25mm
\begin{picture}(300,140)
\put(-00,40){\epsfig{file=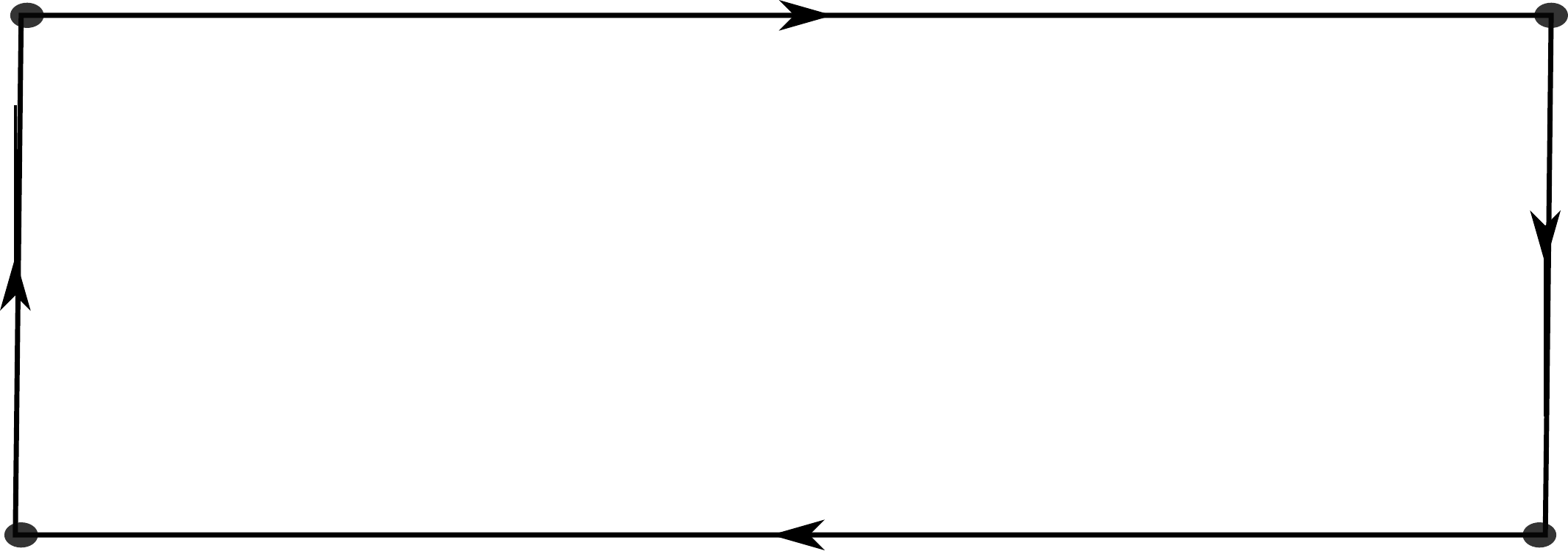,height=40mm, width=80mm}}
\put(80,50){\makebox(0,0)[lc]{$(0,\xi)$}}
\put(80,130){\makebox(0,0)[lc]{$(\infty,\xi)$}}
\put(80,33){\makebox(0,0)[lc]{$\Gamma_3$}}
\put(80,145){\makebox(0,0)[lc]{$\Gamma_1$}} \put(-13,90){\makebox(0,0)[lc]{$\Gamma^-_2$}}
\put(8,90){\makebox(0,0)[lc]{$(\eta,-\infty)$}}
\put(203,90){\makebox(0,0)[lc]{$\Gamma^+_2$}}
\put(160,90){\makebox(0,0)[lc]{$(\eta,+\infty)$}}
\put(-10,145){\makebox(0,0)[lc]{$(\infty,-\infty)$}}\
\put(170,35){\makebox(0,0)[lc]{$(0,+\infty)$}}
\put(-10,35){\makebox(0,0)[lc]{$(0,-\infty)$}}\
\put(170,145){\makebox(0,0)[lc]{$(\infty,+\infty)$}}
\put(-20,15){\makebox(0,0)[lc]{Fig. 2. The domain $\mathfrak{R}$ of definition of the symbol $\mathcal{A}^s_p(\omega)$.}}
\end{picture}

Now we recall the symbol $\mathcal{A}^s_p$ of the operator $\mathbf{A}$ written in [DD16]:
 \begin{subequations}
\begin{equation}\label{e4.2a}
\mathcal{A}^s_p(\omega):=d_0\mathcal{I}^s_p(\omega)
     +\sum_{j=1}^nd_j\mathcal{K}^{1,s}_{c_j,p}(\omega),
     \qquad \omega\in\mathfrak{R}.
\end{equation}
The symbols $\mathcal{I}^s_p(\omega)$ and $\mathcal{K}^{1,s}_{c_j,p}(\omega)$
in \eqref{e4.2a} are defined as follows:
 \begin{eqnarray}\label{e4.2b}
\mathcal{I}^s_p(\omega)&\hskip-3mm:=&\hskip-3mm\begin{cases}
    g^s_p(\infty,\xi), & \omega=(\infty,\xi)\in\overline{\Gamma}_1,
    \\[1ex]
\left(\displaystyle\frac{\eta-\gamma}{\eta+\gamma}\right)^{\mp s}, &
     \omega=(\eta,\pm\infty)\in\Gamma^\pm_2, \\[1ex]
     e^{\pi si}, &\omega==(0,\xi)\in\overline{\Gamma}_3,\end{cases}
 \end{eqnarray}
 \begin{eqnarray}\label{e4.2c}
\mathcal{K}^{1,s}_{c,p}(\omega)&\hskip-3mm:=&\hskip-3mm\begin{cases}
    \displaystyle \frac{e^{-i\pi\left(\frac1p-i\xi-1\right)}c^{\frac1p-i\xi-s-1}}{ \sin\pi(\frac1p-i\xi)},&\omega=(\infty,\xi)\in\overline{\Gamma}_1,\\[1ex]
    0, &\omega=\eta,\pm\infty)\in\Gamma^\pm_2,\\[1ex]
    \displaystyle \frac{e^{-i\pi\left(\frac1p-i\xi-1\right)}c^{\frac1p-i\xi-s-1}}{ \sin\pi(\frac1p-i\xi)},&\omega=(0,\xi)\in\overline{\Gamma}_3,
    \end{cases}\\[1.5ex]
g^s_p(\infty,\xi)&\hskip-3mm:=&\hskip-3mm\frac{e^{2\pi si}+1}2
     -\frac{e^{2\pi si}-1}{2i}\cot\pi\Big(\frac1p-i\xi\Big)=e^{\pi si}\frac{\sin\pi\Big(\frac1p-s-i\xi\Big)}
     {\sin\pi\Big(\frac1p-i\xi\Big)},\quad\xi\in\mathbb{R},\nonumber
 \end{eqnarray}
\end{subequations}
where
 \[
0<\arg\,c<2\pi,\quad-\pi<\arg(c\,\gamma)<0,\quad 0<\arg\gamma<\pi,\quad c^\gamma=|c|^\gamma e^{i\gamma\arg\,c}.
 \]
 %
 \begin{proposition}[\cite{DD16}, Theorem 5.4, \cite{Du15}, Theorem 4.14]\label{p4.1}
Let $1<p<\infty$, $s\in\mathbb{R}$. The operator
 \begin{equation}\label{e4.3}
\mathbf{A}:\widetilde{\mathbb{H}}{}^s_p(\mathbb{R}^+)\longrightarrow
     \mathbb{H}^s_p (\mathbb{R}^+)
 \end{equation}
defined in \eqref{e4.1} is Fredholm if and only if its symbol $\mathcal{A}^s_p(\omega)$ defined in  \eqref{e4.2a}--\eqref{e4.2c}, is elliptic. If $\mathbf{A}$ \, is Fredholm, then
\begin{equation*}
{\rm Ind}\mathbf{A}=-{\rm ind}\det\mathcal{A}^s_p.
\end{equation*}

The operator $\A$. defined in \eqref{e4.1}, is locally invertible at $0$ in the setting \eqref{e4.3} if and only if its symbol $\mathcal{A}^s_p(\omega)$ defined in  \eqref{e4.2a}--\eqref{e4.2c}, is elliptic on $\Gamma_1$:
 \begin{equation}\label{e4.4}
\inf_{\omega\in\Gamma_1}\det\mathcal{A}^s_p(\omega)\not=0.
\end{equation}
\end{proposition}
 %
 \begin{proposition}[\cite{Du15,DD16}]\label{p4.2}
Let $1<p<\infty$, $s\in\mathbb{R}$ and let $\mathbf{A}$ be defined by \eqref{e4.1}. If the operator $\mathbf{A}\;:\;\widetilde{\mathbb{H}}{}^s_p (\mathbb{R}^+) \longrightarrow\mathbb{H}^s_p (\mathbb{R}^+)$ is Fredholm (is invertible) for all $a\in(s_0,s_1)$ and $p\in(p_0,p_1)$, where $-\infty<s_0<s_1 <\infty$,  $1<p_o<p_1<\infty$, then
 \[
\mathbf{A}\;:\;\widetilde{\mathbb{W}}{}^s_p (\mathbb{R}^+) \longrightarrow
     \mathbb{W}^s_p (\mathbb{R}^+),\qquad s\in(s_0,s_1), \quad p\in(p_0,p_1)
 \]
is Fredholm (is invertible, respectively) in the Sobolev-Slobode\v{c}kii  spaces $\mathbb{W}^s_p$ and has the same index
 \[
{\rm Ind}\,\mathbf{A}=-{\rm ind}\,\det\,\mathcal{A}^s_p.
 \]
\end{proposition}
 %
 \begin{proposition}[\cite{Du15,DD16}]\label{p4.3}
Let $1<p<\infty$, $s\in\mathbb{R}$ and let $\mathbf{A}$ be defined by \eqref{e4.1}. If the operator $\mathbf{A}\;:\;\widetilde{\mathbb{H}}{}^s_p (\mathbb{R}^+) \longrightarrow\mathbb{H}^s_p (\mathbb{R}^+)$ is Fredholm (is invertible) for all $a\in(s_0,s_1)$ and $p\in(p_0,p_1)$, where $-\infty<s_0<s_1 <\infty$,  $1<p_o<p_1<\infty$, then
 \[
\mathbf{A}\;:\;\widetilde{\mathbb{W}}{}^s_p (\mathbb{R}^+) \longrightarrow
     \mathbb{W}^s_p (\mathbb{R}^+),\qquad s\in(s_0,s_1), \quad p\in(p_0,p_1)
 \]
is Fredholm (is invertible, respectively) in the Sobolev-Slobode\v{c}kii spaces $\mathbb{W}^s_p$ and has the equal index
 \[
{\rm Ind}\,\mathbf{A}=-{\rm ind}\,\det\,\mathcal{A}^s_p.
 \]
\end{proposition}

\section{Investigation of the Boundary integral equation of model problem}
\label{sect5}
\setcounter{equation}{0}

\noindent
{\bf Proof of Theorem \ref{t0.6}:} Due to Lemma \ref{l3.2} the boundary pseudodifferential equation  \eqref{e0.13} of the model mixed  boundary value problem is Fredholm if the pseudodifferential equation  \eqref{e3.10} is locally invertible at $0$.

Let us investigate the boundary integral equation \eqref{e3.10}. For this it is convenient to rewrite it as an operator equation
\begin{eqnarray}\label{e5.1}
&\M_\alpha\Phi=\F, \\[3mm]
&\Phi:=\left(\begin{array}{c}\vf\\ \psi\end{array}\right)\in
     \widetilde{\bH}{}^r_p(\bR^+),\qquad {\bf
     F}:=\left(\begin{array}{c}G\\
     H\end{array}\right)\in\bH^r_p(\bR^+)\nonumber              \end{eqnarray}
where the operator $\M_\alpha\;:\;\widetilde{\bH}{}^r_p(\bR^+)
\to{\bH}^r_p(\bR^+)$ has the form
\begin{eqnarray*}
\M_\alpha:=\left[\begin{array}{cc} I & -\dst\frac12[\K^1_{e^{i\alpha}}
     +\K^1_{e^{i(2\pi-\alpha)}}]\\
     \dst\frac12[\K^1_{e^{i\alpha}}+\K^1_{e^{i(2\pi-\alpha)}}] & I \end{array}\right]
\end{eqnarray*}
since $\K^1_{e^{i(-\alpha)}}=\K^1_{e^{i(2\pi-\alpha)}}$. Now
Propositions \ref{p4.1} and \ref{p4.2} can be applied to $\M_\alpha$.

We investigate the equation \eqref{e5.1} in the Bessel potential space setting \eqref{e0.15b}. The proof for the Sobolev-Slobode\v{c}kii  spaces $\mathbb{W}^s_p$ follows then from Proposition \ref{p4.2} and we leave the details of the proof to the reader.

Since
\begin{eqnarray*}
&&\hskip-15mm\dst\frac{e^{-i\pi(\Xi-1) + i\alpha(\Xi-r-1)} +
     e^{-i\pi(\Xi-1)+i(2\pi - \alpha)(\Xi-r-1)}}{
     2\sin\pi\Xi}\\[3mm]
&&=e^{-i\pi(\Xi-1)+i\pi(\Xi-r-1)}\dst\frac{e^{i(\pi-\alpha)(\Xi-r-1)}
     +e^{-i(\pi-\alpha)(\Xi-r-1)}}{2\sin\pi\Xi}\\[3mm]
&&=e^{-\pi ri}\dst\frac{\cos[(\pi-\alpha)(\Xi-r-1)]}{\sin\pi\Xi},
 \end{eqnarray*}
using formula  \eqref{e4.2a}-\eqref{e4.2c} we write the symbol of the operator $\M_\alpha$:
\begin{eqnarray}\label{e5.2}
\cM^r_{\alpha,p}(\omega)&\hskip-3mm=&\hskip-3mm\left[\begin{array}{cc} \mathcal{I}^r_p(\omega)
     &-\dst\frac12[\mathcal{K}^{1,r}_{e^{i\alpha},p}+\mathcal{K}^{1,r}_{
     e^{i(2\pi-\alpha)},p}](\omega)\\ \dst\frac1{2}[\mathcal{K}^{1,r}_{e^{i\alpha},p}+\mathcal{K}^{1,r}_{e^{i(2\pi-\alpha)},p}](\omega) & \mathcal{I}^r_p(\omega) \end{array}\right] \nonumber\\
&\hskip-3mm=&\hskip-3mm\left[\begin{array}{cc} e^{\pi ri}
     \dst\frac{\sin\pi(\Xi-r)}{\sin\pi\Xi}&\hskip-7mm -e^{-\pi ri}
     \dst\frac{\cos[(\pi-\alpha)(\Xi-r-1)]}{\sin\pi\Xi}\\[3mm]
     e^{-\pi ri}\dst\frac{\cos[(\pi-\alpha)(\Xi-r-1)]}{\sin\pi\Xi}
     &e^{\pi ri}\dst\frac{\sin\pi(\Xi-r)}{\sin\pi\Xi}
     \end{array}\right],\\[3mm]
&&\hskip25mm\text{for}\quad \omega=(\infty,\xi)\in\overline{\Gamma_1},
     \quad\xi\in\bR,\quad\Xi:=\dst\frac1p-i\xi,\nonumber
 \end{eqnarray}

We did not write the symbol on $\Gamma_2^\pm$ and $\Gamma_3$, because we are only interested in the local invertibility of $\A$ at $0$ (see Theorem \ref{t0.5}, Lemma \ref{l3.1}, Lemma \ref{l3.2} and Proposition \ref{p4.1}).

From \eqref{e5.2} follows:
\begin{eqnarray*}
\det\,\cM^r_{\alpha,p}(\omega)=e^{-2\pi ri}
     \dst\frac{e^{4\pi ri}\sin^2\pi(\Xi-r)+
     \cos^2[ (\pi -\alpha)(\Xi-r-1)]}{\sin^2\pi\Xi},\\
\omega=(\infty,\xi)\in\overline{\Gamma_1}.
 \end{eqnarray*}

Since $e^{4\pi ri}=cos(4\pi r)+i\sin(4\pi r)$, from the latter formula follows that the symbol $\cM^r_{\alpha,p}(\omega)$ {\bf is elliptic} on $\Gamma_1$ if:
{ 
 \begin{eqnarray}\label{e5.3}
&&\hskip-10mm\sin(4\pi r)\sin^2\pi\left(\dst\frac1p-r\right)
     =\sin(4\pi r)\sin^2\pi\left(\dst\frac1p-r-1\right)\not=0
     \quad\text{or}\nonumber\\[3mm]
&&\hskip-10mm\cos(4\pi r)\sin^2\pi\left(\dst\frac1p-r\right) +
     \cos^2\left[(\pi-\alpha)\left(\dst\frac1p - r-1\right)\right]=\\[3mm]
&&\hskip-10mm=\cos(4\pi r)\sin^2\pi\left(\dst\frac1p-r-1\right) +
     \cos^2\left[(\pi-\alpha)\left(\dst\frac1p - r-1\right)\right]\not=0.\nonumber
 \end{eqnarray}}
From \eqref{e5.3} we derive the following conditions of the ellipticity:
{ 
\begin{itemize}
\item[1)]
$r\not=\dst\frac n4$ and $r\not=\dst\frac1p-n-1$, $n=0\pm1,\ldots$, and the condition coincides with \eqref{e0.16a}.
\item[2)]
If $\dst\frac1p-r-1=n$, then $\cos(\pi-\alpha)n\not=0,$ i.e., $(\pi-\alpha)n\not=\dst\frac\pi2+\pi k$, $n=\pm1,\pm2,\ldots$, $k=0,\pm1,\ldots$. This condition coincides with \eqref{e0.16b}.
\item[3)]
If $r=\dst\dst\frac n2$, then
 \[
\sin^2\pi\left(\dst\frac1p-\dst\frac n2-1\right)+\cos^2(\pi-\alpha)
     \left(\dst\frac1p - \dst\frac n2-1\right)\not=0
 \]
and the ellipticity condition is
 \[
\dst\frac1p-\dst\frac n2-1\not=m\quad\text{or}\quad(\pi-\alpha)
     \left(\dst\frac1p - \dst\frac n2-1\right)=(\pi-\alpha)m\not=\frac\pi2(2k+1).
 \]
The condition coincides with  \eqref{e0.16c}.
\item[4)]
If $r=\dst\dst\frac n2-\dst\dst\frac14$, then
\begin{eqnarray*}
&&\sin^2\pi\left(\dst\frac1p-\dst\frac n2 -\dst\frac34\right)
     -\cos^2(\pi-\alpha)\left(\dst\frac1p - \dst\frac n2-\dst\frac34\right)\\
&&=\cos^2\pi\left(\dst\frac1p-\dst\frac n2-\frac14\right)
     -\cos^2(\pi-\alpha)\left(\dst\frac1p - \dst\frac n2-\dst\frac34\right)\not=0.
 \end{eqnarray*}
Then the ellipticity condition is
\begin{eqnarray*}
&&\pi\left(\dst\frac1p-\dst\frac n2-\dst\frac14\right)
 -(\pi-\alpha)\left(\dst\frac1p - \dst\frac n2-\dst\frac34\right)
 =\frac\pi2+\alpha\left(\dst\frac1p - \dst\frac n2-\dst\frac34\right)\not=\pi k\\
&&\text{and}\quad\pi\left(\dst\frac1p-\dst\frac n2-\dst\frac14\right)
 +(\pi-\alpha)\left(\dst\frac1p - \dst\frac n2-\dst\frac34\right)\\
 &&=\pi+2\pi\left(\dst\frac1p-\dst\frac n2\right)
 -\alpha\left(\dst\frac1p - \dst\frac n2-\dst\frac34\right)
 \not=\pi(k+1)
 \end{eqnarray*}
and coincides with \eqref{e0.16d}.
\end{itemize}}

Concerning the unique solvability conditions \eqref{e0.17} of the system \eqref{e0.13} in the non-classical setting \eqref{e0.15b}.

If the conditions \eqref{e0.17} hold, one of the conditions  \eqref{e0.16a}-\eqref{e0.16c} hold as well and, therefore, the symbol of the system \eqref{e0.13} is elliptic. Moreover, for $r=-\dst\frac12$, $p=2$ and arbitrary $0<\alpha<2\pi$  the system \eqref{e0.13} has a unique solution (cf. the concluding assertion of Theorem \ref{t0.5}).  But $r=-\dst\frac12$, $p=2$ and arbitrary $0<\alpha<2\pi$ also satisfy  the conditions \eqref{e0.17} and, due to Proposition \ref{p4.3}  the system of boundary integral equations \eqref{e0.13} has a unique solution for all values of the parameters $\alpha$, $r$ and $p$ which satisfy the  conditions \eqref{e0.17}.   \QED.

\noindent
{\bf Proof of Theorem \ref{t0.3}:} Due to the Theorem \ref{t0.5} the BVP \eqref{e0.4} is Fredholm in the non-classical setting \eqref{e0.6} if the system \eqref{e0.13} in the setting \eqref{e0.15a} is, provided $r=s-1-\dst\frac1p$. The ellipticity condition \eqref{e5.3} for the BVP \eqref{e0.4} acquires the form \eqref{e0.10} and can also be written in the form:
{
 \begin{eqnarray}\label{e5.9}
&&\hskip-10mm\sin\,4\pi\left(s-\dst\frac1p\right)
     \sin^2\pi\left(s-\dst\frac2p\right)\not=0\quad\text{or}\\[2mm]
&&\hskip-10mm\cos\,4\pi\left(s-\dst\frac1p\right)\sin^2\pi
     \left(s-\dst\frac2p\right) +\cos^2(\pi-\alpha)\left(s -\dst\frac2p\right)\not=0.\nonumber
 \end{eqnarray}
From \eqref{e5.3} we get the following conditions of the ellipticity (we have to take into the account the constraint $\dst\frac1p<s<1+\dst\frac1p$):
\begin{itemize}
\item[1)]
$s\not=\dst\frac1p+\dst\frac n4$ and $s\not=\dst\frac2p+n$, $n=0\pm1,\ldots$, and this condition coincides with \eqref{e0.10a}.
\item[2)]
If $s=\dst\frac2p+n$, then $\cos(\pi-\alpha)n\not=0,$ i.e., $(\pi-\alpha)n\not=\dst\frac\pi2+\pi k$, $n=\pm1,\pm2,\ldots$, $k=0,\pm1,\ldots$. This condition coincides with \eqref{e0.10b}.
\item[3)]
If $s=\dst\frac1p+\dst\frac n2$, then
$\sin^2\pi\left(\dst\frac n2-\dst\frac1p\right)+\cos^2(\pi-\alpha)
\left(\dst\frac n2-\dst\frac1p\right)\not=0$, the ellipticity condition is
 \[
\dst\frac n2-\dst\frac1p\not=m\quad\text{or}\quad(\pi-\alpha)
     \left(\dst\frac n2-\dst\frac1p\right) =(\pi-\alpha)m\not=\frac\pi2(2k+1)
 \]
and coincides with  \eqref{e0.10c}.
\item[4)]
If $s=\dst\frac1p+\dst\frac n2+\dst\frac14$, then
\begin{eqnarray*}
&&\sin^2\pi\left(\dst\frac n2 +\dst\frac14-\dst\frac1p\right)
     -\cos^2(\pi-\alpha)\left(\dst\frac n2+\dst\frac14-\dst\frac1p\right)\\
&&=\cos^2\pi\left(\dst\frac n2-\dst\frac14-\dst\frac1p\right)
     -\cos^2(\pi-\alpha)\left(\dst\frac n2+\dst\frac14-\dst\frac1p\right)\not=0
 \end{eqnarray*}
and the ellipticity condition is
\begin{eqnarray*}
&&\pi\left(\dst\frac n2-\dst\frac14-\dst\frac1p\right)
 -(\pi-\alpha)\left(\dst\frac n2+\dst\frac14-\dst\frac1p\right)
 =-\frac\pi2+\alpha\left(\dst\frac n2+\dst\frac14-\dst\frac1p\right) \not=\pi k\\
&&\text{and}\quad\pi\left(\dst\frac n2-\dst\frac14-\dst\frac1p\right)
 +(\pi-\alpha)\left(\dst\frac n2+\dst\frac14-\dst\frac1p\right)\\
 &&=2\pi\left(\dst\frac n2-\frac1p\right)
 -\alpha\left(\dst\frac n2+\dst\frac14-\dst\frac1p\right)
 \not=\pi k
 \end{eqnarray*}
and coincides with \eqref{e0.10d}.
\end{itemize}}

The further proof is similar to the proof of Theorem \ref{t0.6}.
\QED

\vskip10mm
\noindent
{\bf R. Duduchava}, {\em A.Razmadze Mathematical Institute,  Tbilisi State University, Tamarashvili str. 6, Tbilisi 0177, Georgia.} \ {\sf email: RolDud@gmail.com}\\
\\
{\bf M. Tsaava}, {\em A.Razmadze Mathematical Institute,  Tbilisi State University, Tamarashvili str. 6, Tbilisi 0177, Georgia.}\ \ {\sf email: m.caava@yahoo.com}
 \end{document}